\documentclass[useAMS,usenatbib,usegraphicx,useepsfig]{mn2e}

\usepackage[usenames]{color}

\usepackage{amsmath}

\usepackage[utf8]{inputenc}

\newcommand{\msun}{${\rm M_{\sun}}$}

\def\ltsima{$\; \buildrel < \over \sim \;$}
\def\simlt{\lower.5ex\hbox{\ltsima}}
\def\gtsima{$\; \buildrel > \over \sim \;$}
\def\simgt{\lower.5ex\hbox{\gtsima}}
\def\kms{{\rm\,km\,s^{-1}}}
\def\kpc{{\rm\,kpc}}

\def\msun{{\rm\,M_\odot}}


\def\ltsima{$\; \buildrel < \over \sim \;$}
\def\gtsima{$\; \buildrel > \over \sim \;$}

\title[Transverse Velocity of the M31 system]{The Transverse velocity of the Andromeda system, \\
derived from the M31 satellite population}
\author[J.-B. Salomon, R. A. Ibata, B. Famaey, N. F. Martin, G. F. Lewis]{J.-B. Salomon$^{1}$\thanks{E-mail:
jean-baptiste.salomon@astro.unistra.fr}, R. A. Ibata$^{1}$, B. Famaey$^{1}$, N. F. Martin$^{1,2}$, G. F. Lewis$^{3}$\\
$^{1}$Observatoire astronomique de Strasbourg, Universit\'{e} de Strasbourg, CNRS, UMR 7550, 11 rue de l'Universit\'{e}, F-67000 Strasbourg, France\\
$^{2}$Max-Planck-Institut f\"{u}r Astronomie, K\"{o}nigstuhl 17, D-69117 Heidelberg, Germany\\
$^{3}$Sydney Institute for Astronomy, School of Physics, A28, The University of Sydney, NSW, 2006, Australia}
\begin{document}

\date{Accepted 2015 December 4. Received 2015 November 25; in original form 2015 September 13}

\pagerange{\pageref{firstpage}--\pageref{lastpage}} \pubyear{2015}

\maketitle

\label{firstpage}

\begin{abstract}
We present a dynamical measurement of the tangential motion of the Andromeda system, the ensemble consisting of the Andromeda Galaxy (M31) and its satellites. The system is modelled as a structure with cosmologically-motivated velocity dispersion and density profiles, and we show that our method works well when tested using the most massive substructures in high-resolution $\Lambda$ Cold Dark Matter ($\Lambda$CDM) simulations. Applied to the sample of 40 currently-known galaxies of this system, we find a value for the transverse velocity of 164.4 $\pm$ 61.8 $\kms$ ($v_{East}$ = -111.5 $\pm$ 70.2 $\kms$ and $v_{North}$ = 99.4 $\pm$ 60.0 $\kms$), significantly higher than previous estimates of the proper motion of M31 itself. This result has significant implications on estimates of the mass of the Local Group, as well as on its past and future history.
\end{abstract}


\section{Introduction}
\label{sec:Introduction}

Determining accurately the motion of the M31 system with respect to the Milky Way is fundamental for constraining the mass, dynamical history and the future evolution of the Local Group (LG). The line-of-sight velocity of M31 was measured for the first time even before it became clear that the Andromeda Nebula was a galaxy \citep{Slipher1913}, and it was then already shown to be moving towards the Milky Way. Ever since, measuring its proper motion or transverse velocity has proven to be much more complicated, and more than a hundred years after this first line-of-sight velocity measurement, the transverse velocity remains hard to constrain reliably.

In this respect, a first giant leap was made by \citet[hereafter vdMG08]{VanderMarel08} who used the line-of-sight kinematics of M31 satellites as a probe of the global transverse velocity of the M31 system. The key assumption in this work was that the satellites of M31 on average follow the motion of M31 through space. The Heliocentric distances to the M31 satellites were nevertheless not used in that study. This yielded a median transverse velocity of 42 $\kms$ with a one standard deviation confidence interval $\le 56\kms$, consistent with zero at the 1-sigma level. 

Ideally, one would use direct proper motion measurements, based on well-defined point-sources such as water masers whose position with respect to background sources can be measured very accurately. This approach has been used with radio interferometry to measure the proper motion of M33 \citep{Brunthaler05} and that of IC10 \citep{Brunthaler07}. Unfortunately, however, most Local Group galaxies do not contain known masers, and it was only recently that five masers were discovered in M31 \citep{Darling11}. To date, no proper motion for M31 has been published based on these masers, which are located along the star-forming ring. In any case, when such measurements become available, it will be necessary to properly model the annular structure of the star-forming ring, to link the kinematics of the masers to that of the galaxy as a whole.

The proper motion of M31 itself was measured for the first time by \citet{Sohn12} with Hubble Space Telescope (HST) data with 5--7 year time baselines in three fields, using compact galaxies as background reference objects. The surprising result of this study was that the transverse velocity was found to be 17 $\pm$ 17 $\kms$, consistent with a purely radial orbit \citep{VanderMarel12}. 

While being an extremely impressive technical tour-de-force, the \citet{Sohn12} study had to make several assumptions that could have led to a biassed result. The three fields probed in that contribution possess stellar populations with very different kinematics: primarily the extended M31 disk, their so-called ``disk'' field, the stellar halo in their ``spheroid'' field, and the Giant Stellar Stream in their ``stream'' field. \citet{Sohn12} modelled the kinematic behaviour of these components to access the underlying M31 transverse motion. However, it is clear that the kinematics of these populations are currently not fully understood, and this introduces significant model-dependent uncertainties into their result. Their final proper motion value is derived from a weighted average of the data in all three fields, but given the uncertainties, it is the ``spheroid'' field that contributes most significantly to that final value. Inspection of star-counts maps \citep[e.g.][]{Ibata14a} shows that the ``spheroid'' field lies at the edge of an inner shell-like structure, where several kinematic substructures have been identified \citep{Gilbert07}. If the field contains a substantial fraction of stars that do not belong to the spheroid (which is assumed to share the average motion of M31), this could seriously affect any derived proper motion measurement.

Further concerns include colour-dependent point spread function differences between the reference population and the stars of interest, and the variation and degradation of the HST cameras over time.

The astrophysical implications of this measurement are immense, so we judged it to be important to undertake a completely independent estimate of the M31 transverse motion using a different method to \cite{Sohn12}. Furthermore, since we are primarily interested in recovering the past history of the Local Group, we desire to uncover the motion of the {\it M31 system of galaxies} rather that just the motion of M31 itself. After all, the disk galaxy may itself be moving in an orbit within the larger structure.
Given that a large number of new M31 satellite galaxies were discovered as part of the PAndAS survey \citep{McConnachie09,Martin13b}, we embarked on the present project to use these halo tracers together with a new maximum likelihood method to probe the transverse motion of the M31 system.

In Sect.~2, we describe our method in detail, and test it on cosmological simulations of Local Group analogs. The method is then applied to actual observations of the M31 satellite system in Sect.~3 where we present our results. Conclusions are drawn in Sect.~4.

\section{Method}
\subsection{Basic idea}\label{meth1}

We build on the approach devised by \cite{VanderMarel08}, using precise information on the phase-space distribution of the satellite system of M31 to infer the proper motion of the host. The key ingredient which allows us to derive a precise value of the transverse velocity is the use, for the first time, of precise distances for the satellite galaxies of M31 \citep{Conn12} plus a greater number satellites thanks to recent discoveries.

The Heliocentric velocity vector ${\bf v_{sat,i}}$ of each $i^{th}$ satellite galaxy with respect to the Sun can be decomposed into
\begin{equation}\label{base}
{\bf v_{sat,i}} = {\bf v_{M31}} + {\bf v_{pec,i}} - {\bf v_{LSR}} - {\bf v_{pec,\odot}} \, ,
\end{equation}
where ${\bf v_{M31}}$ is the velocity of M31 w.r.t. the Milky Way, ${\bf v_{pec,i}}$ the peculiar velocity of the satellite in the frame of M31, ${\bf v_{LSR}}$ is the circular speed of the Local Standard of Rest (LSR) at the Solar position in the Milky Way, and ${\bf v_{pec,\odot}}$ the peculiar velocity of the Sun w.r.t. the LSR. We use for these values a combination of the peculiar motion derived in \citet{Schonrich10} and of the total tangential motion from \citet{Reid14} using the Galactocentric distance of \citet{Gillessen09}. The line-of-sight component of the velocity of M31 is also taken from \citet{DeVaucouleurs91}.  In the following, we will use a frame centred on the Sun with the $z$-axis pointing towards M31, the $x$-axis pointing to the East, and the $y$-axis pointing to the North. The two unknowns we are searching for are thus $v_{M31x}$ and $v_{M31y}$.

The peculiar velocity ${\bf v_{pec,i}}$ of a satellite galaxy is seen as coming from the equilibrium velocity distribution around the host. The dark matter profile of the host, within which the satellites orbit, is taken to be given by a NFW profile \citep{NFW97} with virial radius $r_{200} = 300 \kpc$, concentration $c = 12$, and virial mass $M_{200} = 1\times10^{12} \msun$, corresponding to mean values of recent results \citep{Watkins10, Fardal13}. We then consider the velocity distribution within this halo to be isotropic, which yields, after Eq. 14 of \cite{Lokas01} :

\begin{equation}\label{sigma}
\begin{split}
\sigma_{r_i}^2 = \frac{1}{2} V_{200}^2 c^2g(c)s(1+cs)^2 \Bigl[ \pi^2 - \ln(cs) - \frac{1}{cs} - \frac{1}{(1+cs)^2}\\
- \frac{6}{1+cs} + \left(1 + \frac{1}{c^2s^2} - \frac{4}{cs} - \frac{2}{1+cs} \right) \times \ln(1+cs) \\
+ 3\ln^2(1+cs) + 6 {\rm Li}_2(-cs)\Bigr] \, ,
\end{split}
\end{equation}
where $V_{200}$ is the circular velocity at the virial radius, $s  = r / r_{200}$ where $r$ is the distance between the centre of the satellite and the centre of M31, $g(c) = 1 / (\ln(1+c) - c/(1+c))$, and ${\rm Li}_2$ is the dilogarithm function.

\begin{figure}
\begin{center}
\includegraphics[angle = 0, viewport = 180 108 750 580, clip, width = \hsize]{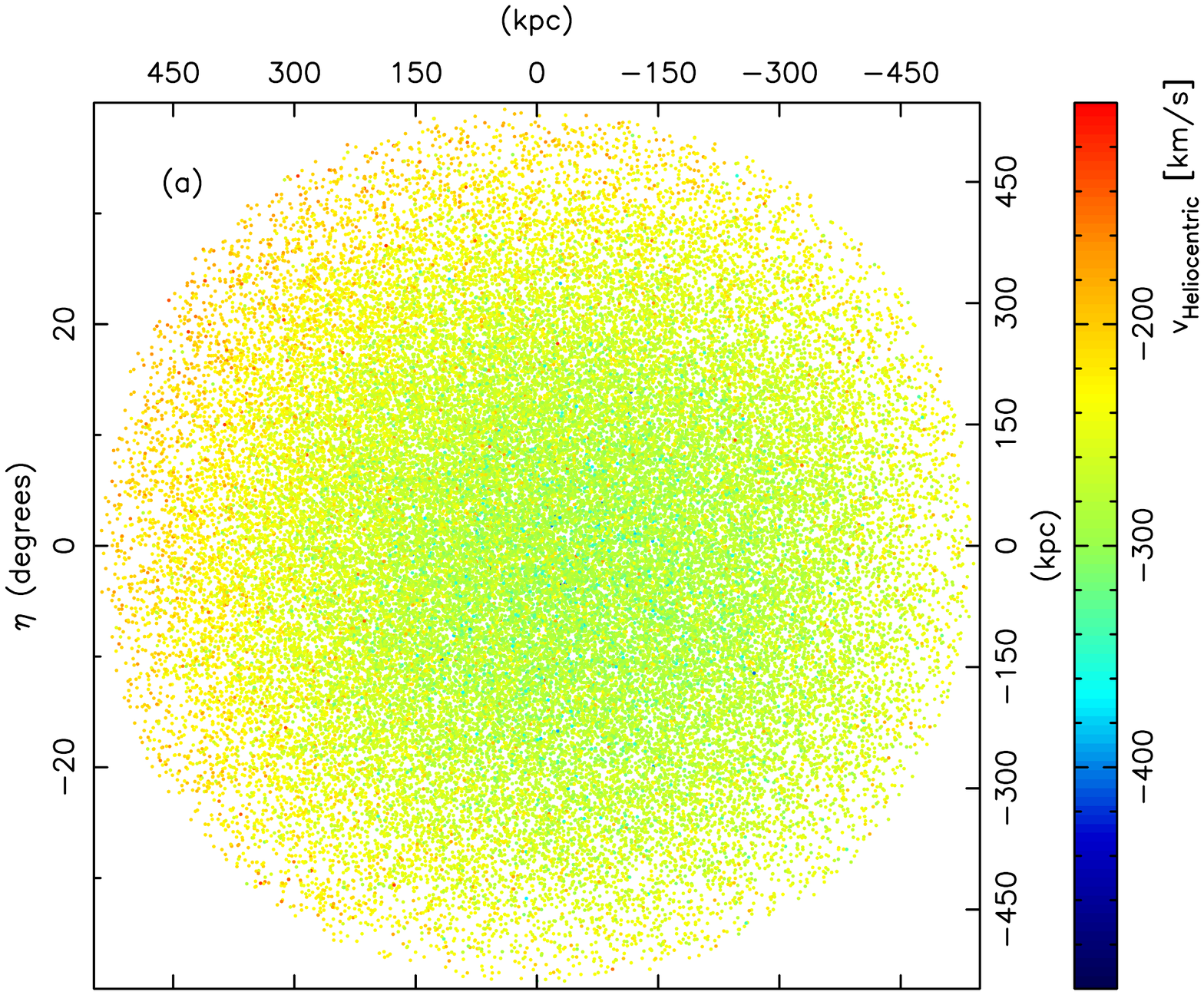} 
\includegraphics[angle = 0, viewport = 180 58 750 542, clip, width = \hsize]{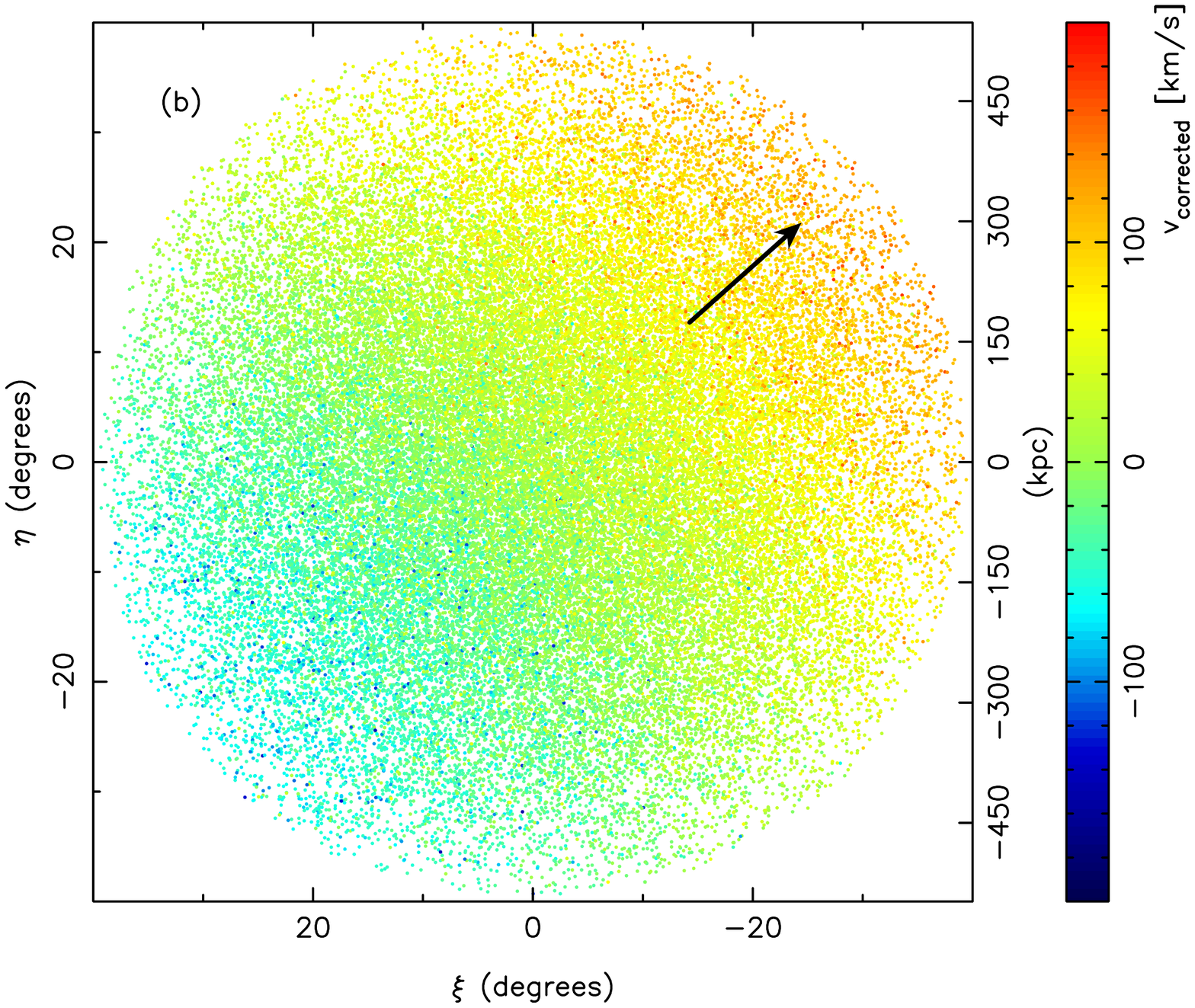}
\caption{\small\label{fig:method}
Illustration of the workings of the method. The simple model described in Section~\ref{mockTest} is used to make a realisation of 50000 satellites which are placed at the distance of M31, and set in motion such that the system possesses the mean motion measured in this contribution. Panel (a) shows the corresponding Heliocentric velocities, which contain a large signal from our motion around the centre of the Milky Way. Once this Solar motion is corrected for (panel b), the bulk motion of the system of satellites becomes apparent as an obvious velocity gradient along the direction of motion of the system (which is indicated by the black arrow). The angular positions (${\xi,\eta}$) are given in terms of standard coordinates with respect to the centre of M31.}
\end{center}
\end{figure}

A MCMC method (typically with $4\times10^{6}$ steps) is used to find the parameters $v_{M31x}$ and $v_{M31y}$ in Eq.~\ref{base} yielding the highest likelihood ($\cal{L}$) weighted by the above dispersion, for the observed line-of-sight velocities and for the given 3D positions of the entire sample of satellites ($n_{sat}$):
\begin{equation}\label{likelihood}
\ln{\cal{L}} = \sum_{i=1}^{n_{sat}} \left[ -\ln({\sigma_{r_i} \sqrt{2\pi}}) - \frac{1}{2} \left(\frac{v_{MCMC_i}-v_{obs_i}}{\sigma_{r_i}} \right)^2 \right],
\end{equation}
with $v_{MCMC_i}$ the MCMC-method predicted velocity on the line of sight for satellite $i$, and $v_{obs_i}$, the observed line-of-sight velocity for the satellite $i$; Figure~\ref{fig:method} provides a visual demonstration of how the method works.

\subsection{Testing the method with cosmological simulations}\label{sec_cosmo}

First of all, we validate our method by applying it to the $z=0$ snapshots of high-resolution cosmological simulations of Local Group analogs undertaken by the ``ELVIS'' collaboration \citep{Garrison-Kimmel14a}. With this simulation, we have at our disposal a zoom on 12 host halo pairs and 24 isolated host halos, all in the mass range of the Milky Way and M31. For each central halo, a large number of particles are accessible beyond the virial radius, where the high resolution allows us to have access to satellites down to a virial mass of $10^8M_{\odot}$. In order to validate our method, we select for each host halo the 39 most massive satellites, i.e. the same number of satellites as in the observational sample of Sect.~3.

\subsubsection{ELVIS : isolated halos}

We first apply the method on the 24 isolated halos in ELVIS. Each of the halos is ``observed" from 10 different points of view, randomly distributed on a sphere of radius 783 kpc, which corresponds to the MW-M31 distance we adopted. This gives a total of 240 ``observed" systems.  We place ourselves in the ``observer" frame, where the $z$-axis is along the line of sight, and the $x$ and $y$ axes arbitrarily chosen while respecting the orthonormality of the coordinate system. For each ``observed" halo, Markov chains of $2 \times 10^6$ steps are built.

\begin{figure}
\begin{center}
\includegraphics[angle = 0, viewport = 95 33 654 562, clip, width = \hsize]{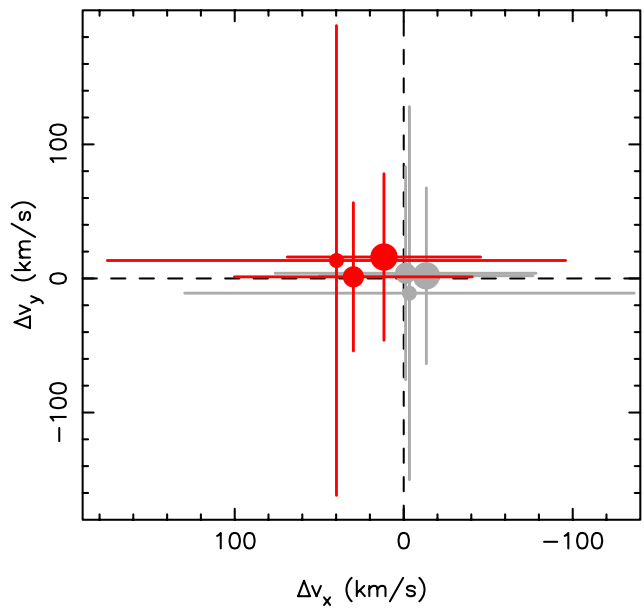}
\caption{\small\label{isigmaVSrvirEvolution2}
Means and uncertainties of the distribution of deviations of the recovered $v_{x}$ and $v_{y}$ w.r.t. the known ones of the ELVIS simulation, using as probes the 39 most massive satellite halos within a given limiting radius $R_{\rm lim}$ in the frame of the host. Grey points correspond to the application of the method to 240 isolated halo ``observations", and red points to 24 halo pair ``observations". The size of the points correspond to the limiting radius: $r_{\rm lim}/r_{200} = 1$ for small points, $r_{\rm lim}/r_{200} = 2$ for medium-sized points, and $r_{\rm lim}/r_{200} = 3$ for the largest points.}
\end{center}
\end{figure}

We plot on Figure~\ref{isigmaVSrvirEvolution2} the mean deviation of our recovered $v_{x}$ and $v_{y}$ parameters with respect to the known relative velocity of the isolated host w.r.t. the observer in the simulation (grey points). Different limits are used for the maximum distance from the host center, $R_{\rm lim}$, of the 39 satellites. The small, medium and large grey points correspond respectively to $r_{\rm lim}/r_{200} = 1$, $r_{\rm lim}/r_{200} = 2$ and $r_{\rm lim}/r_{200} = 3$ where $r_{200}$ is the virial radius of the host halo. 

The transverse velocity is better estimated for $r_{\rm lim}/r_{200} = 2$ than for $r_{\rm lim}/r_{200} = 1$. This is due to the fact that the method needs to sample a large range of lines of sight to be effective. However, this improvement stops being useful when the satellites are too far away from the host because they are not bound to it anymore. The optimum recovery is for $r_{\rm lim}/r_{200} \sim 2$ with a distribution of deviations centred on the correct value to better than 10 $\kms$ with 65 $\kms$ uncertainties in both directions.

\subsubsection{ELVIS: halo pairs}

We now apply the same procedure to the 12 halo pairs of ELVIS. This time, the ``observation" is made from the point of view of an observer located at the centre of the other halo of the pair. This again leads to 24 ``observations" of satellite systems. In Figure~\ref{isigmaVSrvirEvolution2} the mean deviation of our recovered $v_{x}$ and $v_{y}$ with respect to the true relative velocity is plotted in red. The optimum recovery is also attained for $r_{\rm lim}/r_{200} \sim 2$, with a distribution of deviations centred on $15\kms$ and $\sigma=55\kms$ in both directions. This slightly smaller $\sigma$ than in the isolated case is probably due to the limited influence of the environment in the halo pair case. The gravitational effect of the environment is dominated by the second halo in the pair configurations, while for the isolated halos it varies from case to case.

In summary, the performance of the proposed method is validated by these high-resolution cosmological simulations, and the associated typical error is very reasonable ($\sim 55\kms$) with respect to previous studies that used satellites \citep{VanderMarel08} to constrain the M31 proper motion. Moreover, these tests show that our method, which fits a spherically-symmetric halo model, works well even when applied to realistic triaxial haloes.

\subsection{Testing the method with mock M31 satellite systems}\label{mockTest}

For our observational study, all M31 satellites known to date will be considered: And I, And II, And III, And V, And VI, And VII, And IX, And X, And XI, And XIII, And XV, And XVI, And XVII, And XVIII, And XIX, And XX, And XXI, And XXII, And XXIII, And XXIV, And XXV, And XXVI, And XXVIII, And XXIX, And XXX (Cass II), NGC 147, NGC 185, M32, NGC2 05, IC 10, LGS 3, Pegasus, IC 1613, M33, Per I, Lac I, Cass III, And XII and And XIV. 

The positions on the sky and the line-of-sight velocities of these satellites have been extracted from the recent literature \citep{McConnachie12, Collins13, Martin14}. The distances of 11 satellites are also extracted from the same papers, while that of the 25 other satellites have been estimated by \cite{Conn12}. The distances of Lac I, and Cass III are taken from \cite{Martin13a} and that of Per I from \cite{Martin13c}. As explained in Section~\ref{meth1}, we will work in a frame centred on the Sun,  with the $z$-axis pointing towards the centre of M31, the $x$-axis pointing to the East, and the $y$-axis pointing to the North.

In order to further validate the method, we construct a simple three-dimensional model representing the M31 satellite system observed from a star orbiting in a neighbouring large spiral galaxy. We put 39 satellites in the mock M31 system exactly at the same position w.r.t. the host as in the observed case. The observer's galaxy is placed at a distance of 783~kpc, with the observer orbiting this galaxy at 8~kpc from its center, at a circular velocity of $220\kms$ and an (arbitrary) peculiar velocity $(U, V, W)_{\odot} = (10,10,10)\kms$. We then impose the relative velocity between the two large galaxies to be (i) ${\bf v_{M31}} = (-100, 100, 100)\kms$, and (ii) ${\bf v_{M31}} = (0, 0, 100)\kms$ to check that a pure radial motion can be recovered by the method. The 39 satellites are located at the mean observed distances from the real M31 system, and a peculiar velocity is randomly drawn in 1000 models from a NFW profile with Eq.~\ref{sigma}. First we use the same halo parameters as in the following MCMC fit, i.e. $c = 12$, $r_{200} = 300\kpc$ and $M_{200} = 1\times10^{12} \msun$.
As expected, we then recover in both cases (i) and (ii) the transverse component of ${\bf v_{M31}}$ with a typical maximum deviation of $\Delta v_{M31x} = 2.4 \pm 73\kms$ and $\Delta v_{M31y} = 0.1 \pm 60\kms$, quite close to the values found in our tests with the ELVIS simulations.
This demonstrates that, even with a sparse spatial sampling of only 39 satellites (with only one {\it spatial} realisation compared to the multiple realisations in ELVIS), the method recovers the true transverse velocity with no bias and a reasonable uncertainty.

Then, we check the robustness of the results to a mistaken choice of velocity distribution in Eq.~\ref{sigma}. For this, we construct models with $M_{200}$ varying from $0.7$ to $2.0\times10^{12}\msun$, concentrations from $8.0$ to $20.0$, and virial radii from $200$ to $350$~kpc. We then apply our MCMC method with the likelihood weighted by the dispersion coming from Eq.~\ref{sigma} with $c = 12$, $r_{200} = 300 \kpc$ and $M_{200} = 1\times10^{12} \msun$. The recovered M31 velocity only typically deviates by $0.3 \kms$ from the true ones, with the same dispersion as before. This means that an inadequate choice of halo parameters has a negligible impact on the uncertainties.

We also check the influence of distance uncertainties. We apply our method by using distances drawn from the observational PDF (from \cite{Conn12} for 25 satellites and from Gaussian PDF built from the observational uncertainties for the others) at each step in our Markov Chains. The typical deviations induced by these uncertainties is only $0.6{\pm0.1} \kms$ on $\Delta v_{M31x}$ and $1.5{\pm2.7} \kms$ on $\Delta v_{M31y}$. We will use this method based on the PDF for the application of the method to the true observations in the next section.

\begin{table*}
 \centering
  \caption{Observational results in terms of Heliocentric velocities, corrected for the LSR rotation and the peculiar motion of the Sun.} \label{tabObservations}
  \begin{tabular}{@{}lrrrr@{}}
  \hline
  Method and selections &  East                          &  North                        &      Sky          & Radial           \\
                                           & $ v_{M31x} $           &  $ v_{M31y}$           & $v_{t}$  & $ v_{M31z}$  \\
                                           &    ($\kms$)                     &     ($\kms$)                   &     ($\kms$)       &     ($\kms$)         \\
  \hline
39 satellites, $ v_{M31z}$ fixed  &  $-121.0 {\pm69.6}$ &  $ 80.1 {\pm57.8}$ & $ 159.5 {\pm62.4}$      & $-103.9  {\pm4.0}$\\
39 satellites, $ v_{M31z}$ free   &  $-111.1 {\pm70.0}$ &  $ 100.3 {\pm59.8}$ & $ 164.5 {\pm61.8}$    & $-87.0  {\pm14.0}$\\
\hline
{\bf 40 satellites (M31 as a ``satellite")} &   ${\bf-111.5 {\pm70.2}}$&  ${\bf 99.4 {\pm60.0}}$ & ${\bf 164.4 {\pm61.8}}$     & ${\bf-87.5  {\pm13.8}}$\\
\hline
26 satellites (without VTP)           &  $-77.6 {\pm 70.7}$ &  $ 1.2 {\pm 64.2}$ & $ 109.9 {\pm 55.2}$     & $-84.0  {\pm 16.8}$\\
30 satellites (velocity limitation) &   $-171.2 {\pm79.5}$ &  $ 96.6 {\pm68.3}$ & $ 211.3 {\pm 70.5}$     & $-98.3  {\pm15.4}$\\
\hline
\end{tabular}
\end{table*}

\section{Results}

\subsection{Complete sample of satellites}

We use the MCMC method described in the previous section, with $4\times10^{6}$ steps, in order to find the parameters $v_{M31x}$ and $v_{M31y}$ from Eq.~\ref{base} yielding the highest likelihood weighted by the dispersion of Eq.~\ref{sigma} as in Eq.~\ref{likelihood}, for the observed line-of-sight velocities and for the given 3D positions of the satellites. What is used is actually the full PDF for the distances, taking into account the uncertainties. Each Markov chain will then begin with a random drawing among each of these distance PDFs. We use the halo parameters presented in Sect.~\ref{meth1} to obtain the $\sigma_{r_i}$, using Eq.~\ref{sigma}.
We correct the peculiar velocities for the motion of the Sun using $U_{\odot} = 11.1^{+0.69}_{-0.75}  \kms$, $V_{\odot} + V_{\rm LSR} = 255.2 \pm 5.1 \kms$ and $W_{\odot} = 7.25^{+0.37}_{-0.36} \kms$ \citep{Schonrich10, Reid14}.
Note that we draw a value for $(U_{\odot}, V_{\odot}+ V_{\rm LSR}, W_{\odot})$ at each step of our Markov chain in order to take into account the uncertainties on these parameters.
Finally, we force the radial velocity of the M31 system, $v_{M31z}$, to be the measured radial velocity of the M31 galaxy itself, corrected for the Solar velocity in that direction, i.e. $-103.9  {\pm4.0}\kms$. This prior is given as a gaussian PDF applied to each step of the Markov chain.

The first line of Table~\ref{tabObservations} yields the resulting corrected velocities: $v_{M31x}$ = -121.0 $\pm$ 69.6 $\kms$, $v_{M31y}$ = 80.1 $\pm$ 57.8 $\kms$, $v_{t}$ = 159.5 $\pm$ 62.4 $\kms$ (modulus of the transverse component from $v_{M31x}$ and $v_{M31y}$) and $v_{M31z}$ = -103.9 $\pm$ 4.0 $\kms$, i.e. the East, North, transverse and radial components respectively. These values are Galactocentric, i.e. they give the relative motion of the M31 system centre to the MW centre. The Heliocentric velocities, observed from the Sun, are respectively $23.6 {\pm69.6}\kms$, $0.9 {\pm57.8}\kms$, $82.6 {\pm43.8}\kms$ and $-300.0  {\pm4.0}\kms$. Note that the typical uncertainties yielded by the MCMC method are of the same order of magnitude as the typical dispersion of the results with respect to the true values in simulations, demonstrating the coherence of the method. These uncertainties are also very similar to those obtained by \citet{VanderMarel08} when applying a different method to the M31 system. In their study, the smaller errors they quote result from measuring a weighted average mean to the results obtained by applying their method to different individual objects such as M33 and objects supposedly outside of the Local Group.

\begin{figure}
\begin{center}
\includegraphics[angle = 0, viewport = 190 108 735 580, clip, width = 8cm]{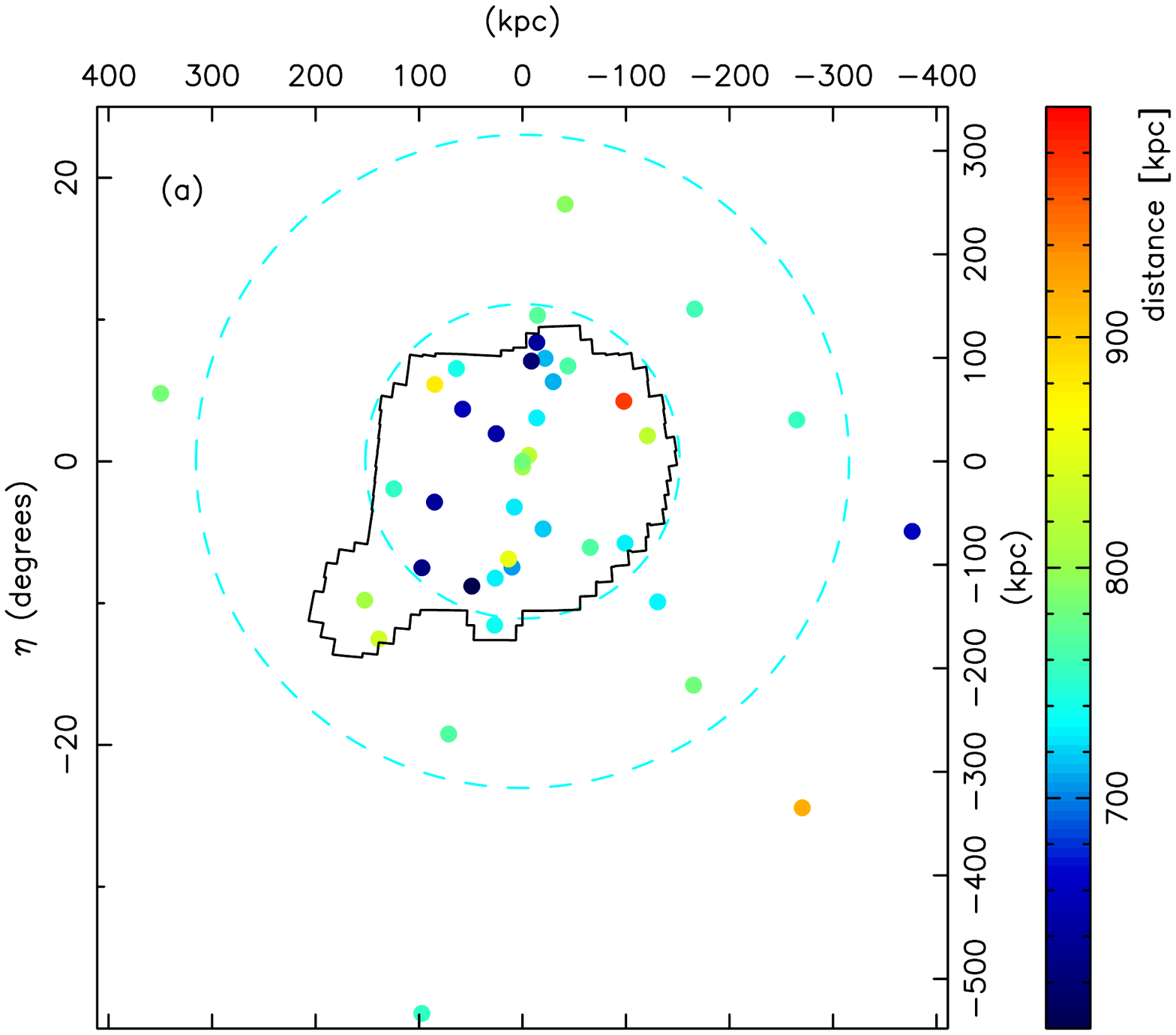} 
\includegraphics[angle = 0, viewport = 190 108 735 539, clip, width = 8cm]{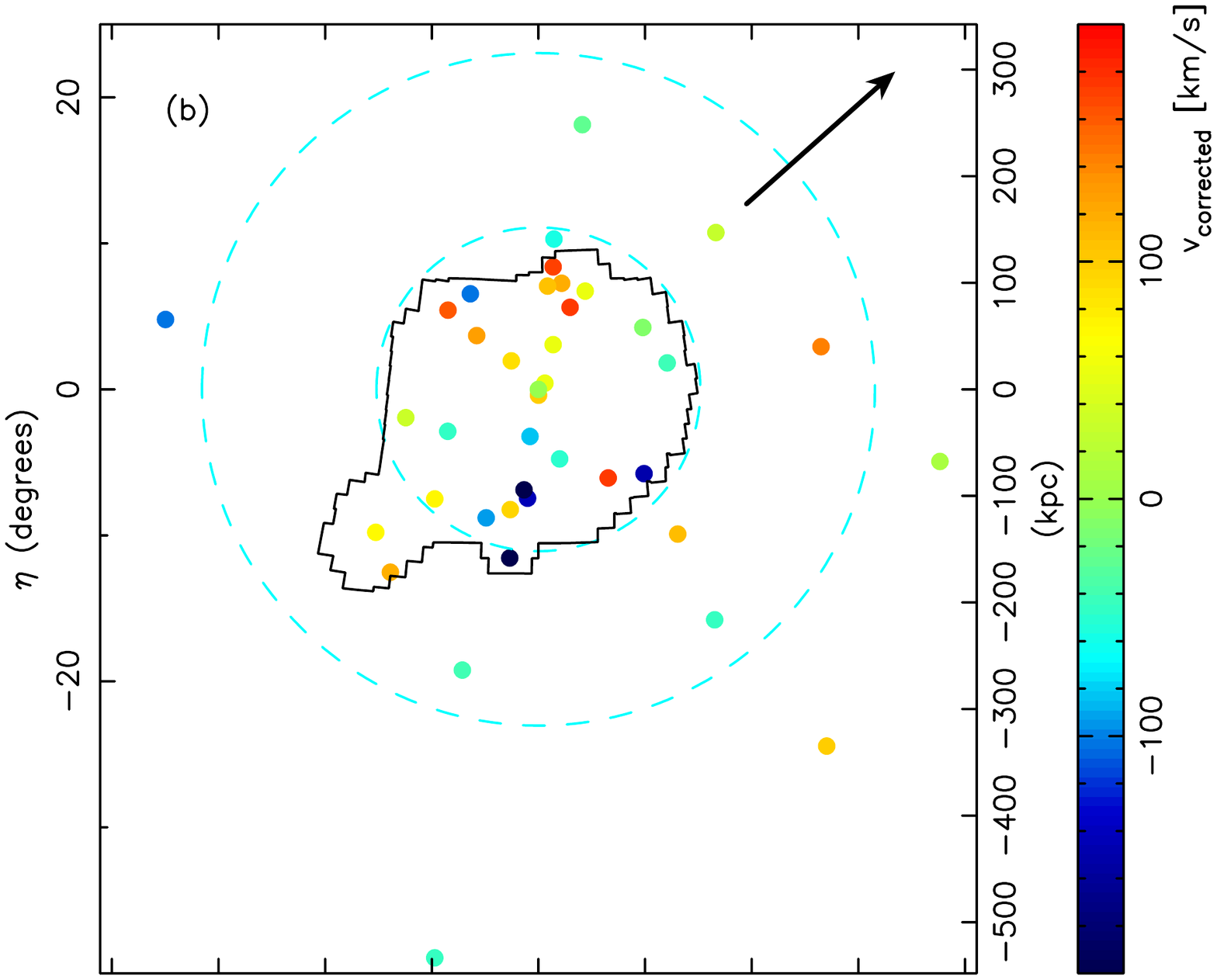}
\includegraphics[angle = 0, viewport = 190 60 735 539, clip, width = 8cm]{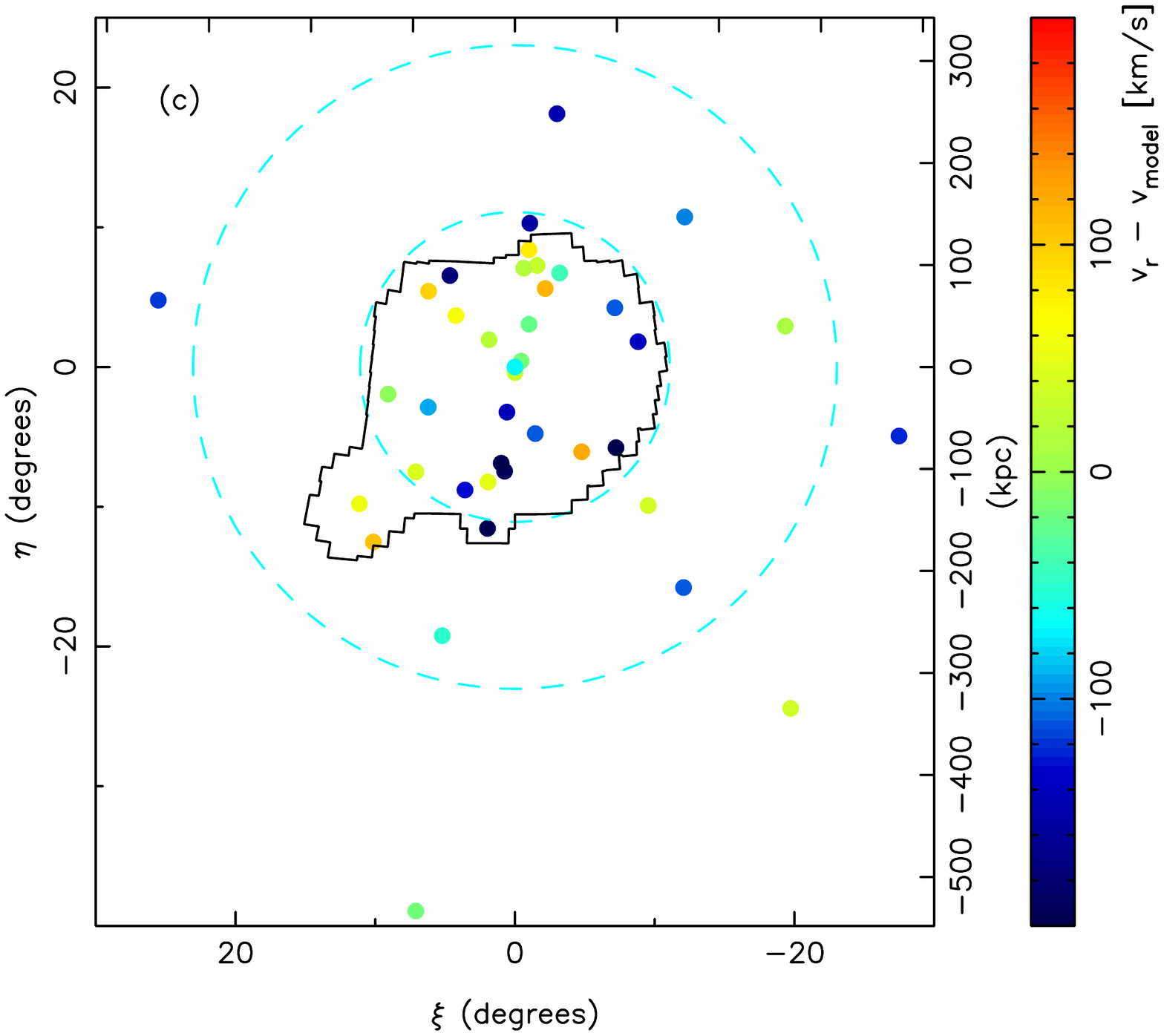}
\caption{\small\label{4a}
Properties of the real satellite sample: (a) the most likely Heliocentric distances, (b) the radial velocity corrected for the Solar motion (compare to Figure~\ref{fig:method}b), (c) the residuals with respect to the best-fit model. The most robust direction of motion recovered by our method is shown with an arrow in (b); while not completely straightforward to interpret visually due to the multi-dimensionality of the information, it can be seen to correspond to a direction along which the velocity gradient is high. The dashed-line circles mark $150\kpc$ and $300\kpc$ ($\approx r_{200}$), while the irregular polygon delineates the boundary of the PAndAS survey. (${\xi,\eta}$) are standard coordinates centred on M31.}
\end{center}
\end{figure}

\subsection{Relaxing the radial velocity constraint}
In a second implementation, we relaxed the constraint on the radial velocity $v_{M31z}$, which is now considered fully as a third unknown in Eq.~\ref{base}. Line 2 of Table~\ref{tabObservations} gives the recovered Galactocentric velocities ($v_{M31x}$ = -111.0 $\pm$ 70.0 $\kms$, $v_{M31y}$ = 100.3 $\pm$ 59.8 $\kms$, $v_{t}$ = 164.5 $\pm$ 61.8 $\kms$ and $v_{M31z}$ = -87.0 $\pm$ 14.0 $\kms$). The Heliocentric values are: $ v_{M31x} =  33.5 {\pm70.1}\kms$, $ v_{M31y} = 20.6 {\pm59.8}\kms$, $v_{t} = 88.7 {\pm46.5}\kms$ and $ v_{M31z} = -282.1  {\pm14.4}\kms$. The one sigma uncertainty of the obtained radial velocity component and of the observed value are overlapping, which provides strong evidence that M31 resides at the centre of its satellite system.

\subsection{Andromeda considered as yet another ``satellite" in the halo satellite system}

In a third implementation, which we consider as the most robust, we add M31 itself as an additional ``satellite" in the halo with radial velocity -103.9 $\pm$ 4.0 $\kms$. The MCMC method is thus now applied to a system of 40 satellites. The distance of each satellite from the Sun is drawn at each step from the observational PDF on the distance. Then the distance to the M31 galaxy itself from the centre of the system is drawn from a gaussian PDF with 25~kpc dispersion, and the distance from each galaxy to the centre of the M31 system is then recomputed. In the case of the M31 galaxy, the distance to itself is obviously zero with no error, and the distance from the centre of the system is just drawn from a gaussian PDF with $25\kpc$ dispersion.

The resulting transverse and radial velocities of the M31 system are indicated in the 3rd line of Table~\ref{tabObservations} ($v_{M31x}$ = -111.5 $\pm$ 70.2 $\kms$, $v_{M31y}$ = 99.4 $\pm$ 60.0 $\kms$, $v_{t}$ = 164.4 $\pm$ 61.8 $\kms$ and $v_{M31z}$ = -87.5 $\pm$ 13.8 $\kms$).
The Heliocentric values are : $ v_{M31x} =  33.1 {\pm70.2}\kms$, $ v_{M31y} = 19.6 {\pm60.0}\kms$, $v_{t} = 88.6 {\pm46.5}\kms$, and $ v_{M31z} = -282.6  {\pm14.3}\kms$.

The velocities are very similar to the ones obtained with 39 satellites, and the error on the radial velocity of the whole M31 system overlaps at the one sigma level with the velocity of the M31 ``satellite". This value is the least model-dependent one, and is based on the largest sample, meaning that the Galactocentric transverse velocity of $164.4 {\pm 61.8}\kms$ is considered as our best estimate.

\begin{figure}
\begin{center}
\includegraphics[angle = 0, viewport = 95 33 654 562, clip, width = \hsize]{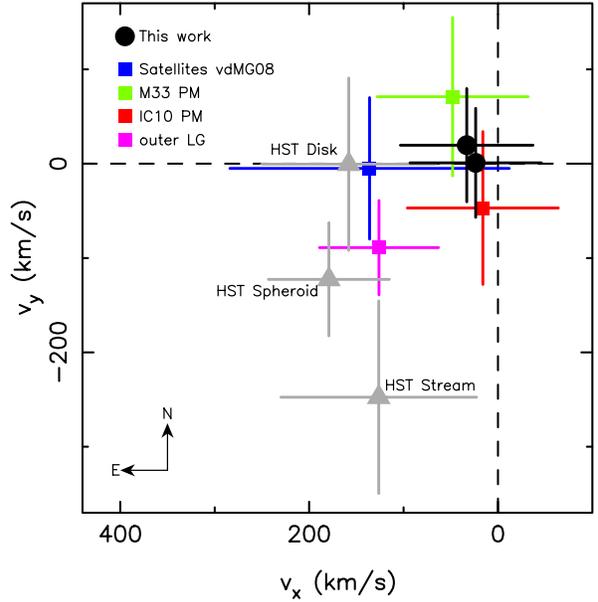}
\caption{\small\label{resultsHelio}
Heliocentric transverse velocities of the M31 system and M31 itself. The results of the present study for 39 satellites with $v_{M31z}$ fixed, and for 40 satellites with $v_{M31z}$ as a free parameter are shown with large black circles. For comparison, we also show the different estimations of the Heliocentric transverse motion of the M31 complex measured by vdMG08 with squares of different colours. The vdMG08 satellite sample is shown in blue, the constraint from M33 in green, the constraint from IC10 in red, and the constraint from the outer Local Group sample in magenta. The proper motion measurements of M31 stars from the three deep HST fields studied by \citet{VanderMarel12} are shown in grey. These three HST field values are shown shifted to reflect the M31 centre of mass motion, i.e. the HST PMs are corrected for the internal kinematics model and the viewing perspective.}
\end{center}
\end{figure}

\subsection{Influence of the plane of satellites}\label{plan}

It has recently been claimed that approximately $50$\% of the satellite galaxies of M31 are confined to a Vast Thin Plane (VTP) of satellites \citep{Ibata13}. The presence of this structure could in principle bias our transverse motion estimates, so we decided to also apply the method to the 26 satellites that are not part of the VTP. The recovered values are listed in line 4 of Table~\ref{tabObservations} ($v_{M31x}$ = -77.6 $\pm$ 70.7 $\kms$, $v_{M31y}$ = 1.2 $\pm$ 64.2 $\kms$, $v_{t}$ = 109.9 $\pm$ 55.2 $\kms$ and $v_{M31z}$ = -87.5 $\pm$ 13.8 $\kms$), corresponding to Heliocentric values of $ v_{M31x} =  67.0 {\pm 70.7} \kms$, $ v_{M31y} = -78.5 {\pm 64.2} \kms$, $v_{t} = 128.0 {\pm 58.2} \kms$ et $ v_{M31z} = -279.1  {\pm 17.2} \kms$. The one sigma errors from this and the previous estimations based on 39 and 40 objects are overlapping. Since the VTP has a North-South orientation, a lot of information has nevertheless been lost, and the North component $v_{M31y}$ has been the most affected.

Nevertheless, to be sure that a co-rotating plane cannot substantially affect the results, we have applied our method on satellite planes selected from the ELVIS simulations.
Once again, the 24 isolated haloes are ``observed'' from 10 different MW-like points of view.
We first define an edge-on plane of satellites, by selecting the set of 13 satellite haloes that have velocities most consistent with being in co-rotation with thickness $<20\kpc$ (but no mass criterion is used for the selection). The simulation is then rotated so that the plane is seen edge-on in the North-South direction to the fictitious observer.
To obtain a full sample of 39 satellites, we complement the co-rotating sample by adding in the 26 most massive satellites from the simulation.
This selection procedure therefore produces systems with an apparent co-rotating plane of satellites similar to that of M31, and with coherent dynamics.
Note that the maximum distance of satellites from the host is always limited to 2 virial radii.
We applied the method to the 240 observations.
The mean deviation of our recovered $v_{x}$ and $v_{y}$ parameters with respect to the known relative velocity of the isolated host is $\Delta v_{x} = -4.0 {\pm 89.6} \kms$ and $\Delta v_{y} = 52.1{\pm 87.6} \kms$.
The uncertainties are $\sim 15 \kms$ larger than those obtained in Section \ref{sec_cosmo}, which seems reasonable given that we have forced a correlation in the data.
In the direction perpendicular to the plane ($v_{x}$), we recover the correct velocity, yet even in the direction parallel to the edge-on  plane ($v_{y}$), our strict selection has only a limited impact since the systematic deviation is smaller than the uncertainties.
Consequently, we see that even with 13 satellites in a 3D co-rotating plane, the method is robust enough to find the tangential velocity of the host halo.

\subsection{Influence of the satellites with high velocities}

It was shown previously in section \ref{mockTest} that the assumed velocity distribution has little impact on the obtained result even when wrong values for the velocity distribution parameters in Eq.~\ref{sigma} are used. However, the true halo mass has another effect, namely its capacity to bind satellites. Given a certain halo profile, the speed of a satellite can be too high for it to be bound to the system, namely when it exceeds the escape speed
\begin{equation}
v_{esc} = \sqrt{\frac{2 G M_{200}}{r}} \, .
\end{equation}
This can be the case for satellites that have just arrived and are passing by, or if they have undergone strong interactions involving a gravitational kick. In such a situation, the method clearly sees the given line-of-sight velocity of the satellite as an exception and attempts to minimize the deviation. Previous tests on cosmological simulations have shown that this was, in principle, not a big issue, mainly because such exceptions tend to compensate each other. We nevertheless decided to make a further measurement, restricting our sample to satellites with
\begin{equation}\label{vesc}
v_{esc} > \sqrt{3} \,  | v_{los}-v_{M31z} |  \, .
\end{equation}
following the criterion of \cite{McConnachie12} where the escape speed is calculated for the NFW model used in the MCMC method.
This criterion limits our sample to 30 satellites where And XIX, And XX, And XXII, And XXIV, And XXV, AndXXX, Pegasus, M33, And XII and And XIV are excluded.
The mock M31 model used in the previous section is used again with 30 satellites, and it is then found that the associated sparse sampling induces a systematic deviation from the true transverse velocity, $\Delta v_{M31x} = 38.6 {\pm 50.1} \kms$, $\Delta v_{M31y} = -12.5 {\pm 41.5} \kms$ and $\Delta v_{M31z} = -2.5 {\pm 11.8} \kms$. This is because the observational criterion to exclude galaxies depends on the line of sight, and on the actual orientation of the velocity vector of the satellite. We subtract theses systematic deviations from our obtained results, and report the final values in the final line of Table~\ref{tabObservations} ($v_{M31x}$ = -171.2 $\pm$ 79.5 $\kms$, $v_{M31y}$ = 96.6 $\pm$ 68.3 $\kms$, $v_{t}$ = 211.3 $\pm$ 70.5 $\kms$ and $v_{M31z}$ = -98.3 $\pm$ 15.4 $\kms$). The Heliocentric values are: $ v_{M31x} =  -26.6 {\pm 79.5} \kms$, $ v_{M31y} = 16.9 {\pm 68.3} \kms$, $v_{t} = 96.9 {\pm 50.8} \kms$ and $ v_{M31z} = -293.3  {\pm 15.8} \kms$. Because of the reported systematic deviations that we had to subtract from our results, we consider this case less reliable than the previous ones. Indeed, a slight increase of the mass of the halo would allow almost all satellites to be bound, thus getting back to the previous cases,  without any systematic bias on the result.

The effect of outliers in a sample can also
be assessed in an automatic way \citep{Sivia06}.
To this end, we adopted the ``conservative formulation'' of \cite{Sivia06},
which involves a modification of the likelihood equation (Eq. \ref{likelihood}), where the contribution 
of outliers are marginalized in the calculation:
\begin{equation}\label{likelihood2}
\begin{split}
\ln{\cal{L}} =  \sum_{i=1}^{n_{sat}} \Bigr\{  -\ln({\sigma_{r_i} \sqrt{2\pi}}) + \ln \left[ 1 - e^ {\left( - T_i^2 / 2 \right)  } \right] 
- \ln \left( T_i^2 \right) \Bigl\} \, ,
\end{split}
\end{equation}
where $T_i = (v_{MCMC_i}-v_{obs_i}) / \sigma_{r_i}$.
We first applied this method to the satellites of the 24 ELVIS isolated halos observed from 10 different MW-like points of view (as previously selected in Section \ref{sec_cosmo}).
For the 2 parameters of interest $v_{x}$ and $v_{y}$, we obtained the same central values.
The uncertainties are about 5 $\kms$ larger than those calculated with the standard likelihood formulation (Equation \ref{likelihood}).
Then, to compare the two methods on the observations, we applied the ``conservative formulation'' to the complete sample of 40 satellites where the third parameter ($v_z$) is also relaxed.
The Heliocentric values are: $ v_{M31x} =  55.1 {\pm 133.6}\kms$, $ v_{M31y} = 12.6 {\pm 107.8}\kms$, $v_{t} = 156.8 {\pm 89.8}\kms$ and $ v_{M31z} = -277.1  {\pm 24.8}\kms$ (corresponding to Galactocentric velocities of: $ v_{M31x} =  -89.5 {\pm 133.6}\kms$, $ v_{M31y} = 92.3 {\pm 107.8}\kms$, $v_{t} = 191.9 {\pm 95.6}\kms$ and $ v_{M31z} = -82.0  {\pm 24.6}\kms$).
These values are in good agreement with the values calculated with the standard likelihood formulation (which are listed in the 3$^{\rm rd}$ line of Table~\ref{tabObservations}).
Thus, the central values obtained both on cosmological simulations and observations are consistent, which proves that our approach is robust to the presence of outliers.
Nevertheless, the uncertainties are significantly larger with the ``conservative formulation'', which is what is expected with this method \citep{Sivia06}, as it effectively reduces the information content of the data.

\subsection{Summary of results}

\begin{figure*}
\begin{center}
\includegraphics[scale=0.315]{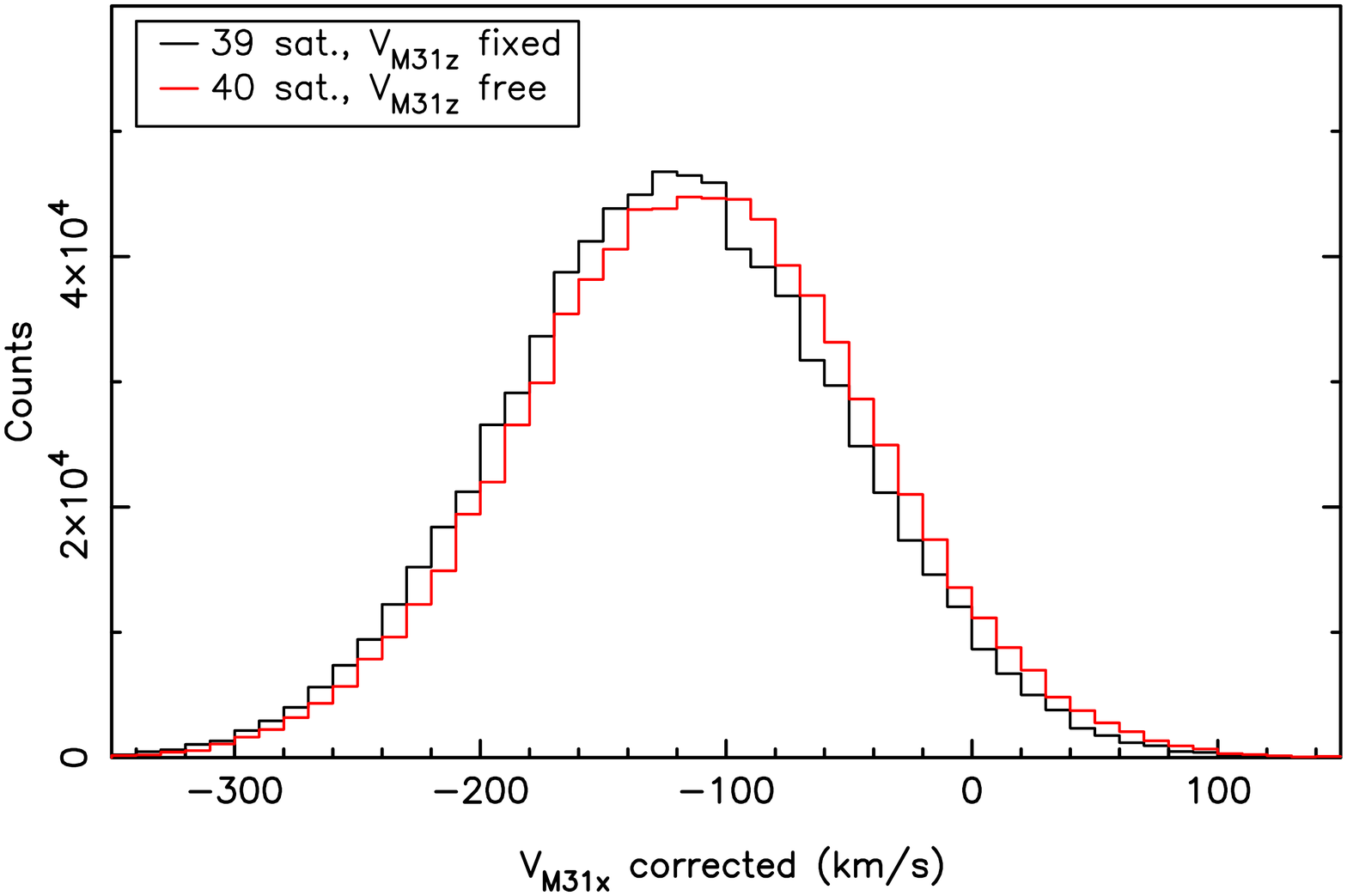} \includegraphics[scale=0.315]{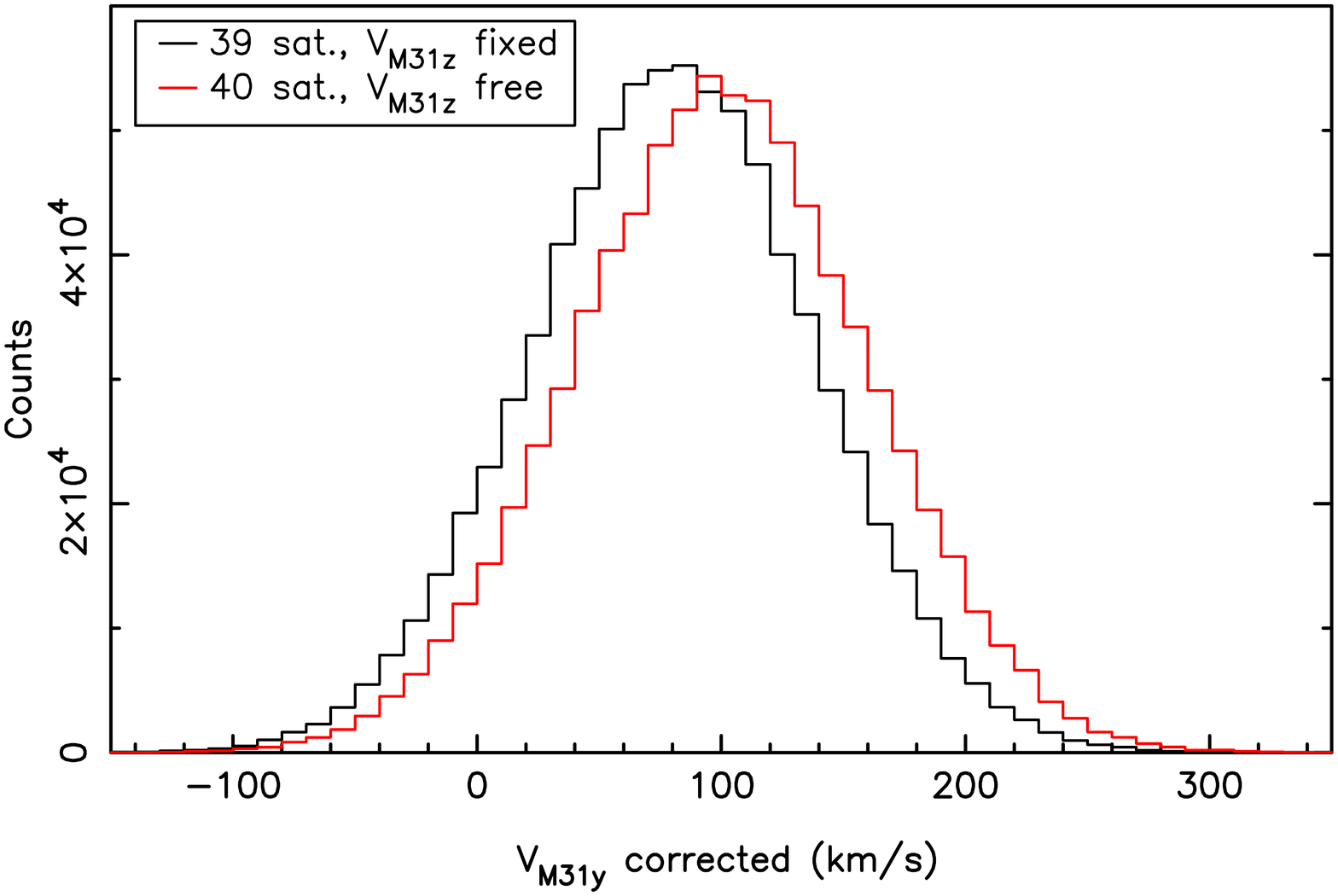}
\includegraphics[scale=0.315]{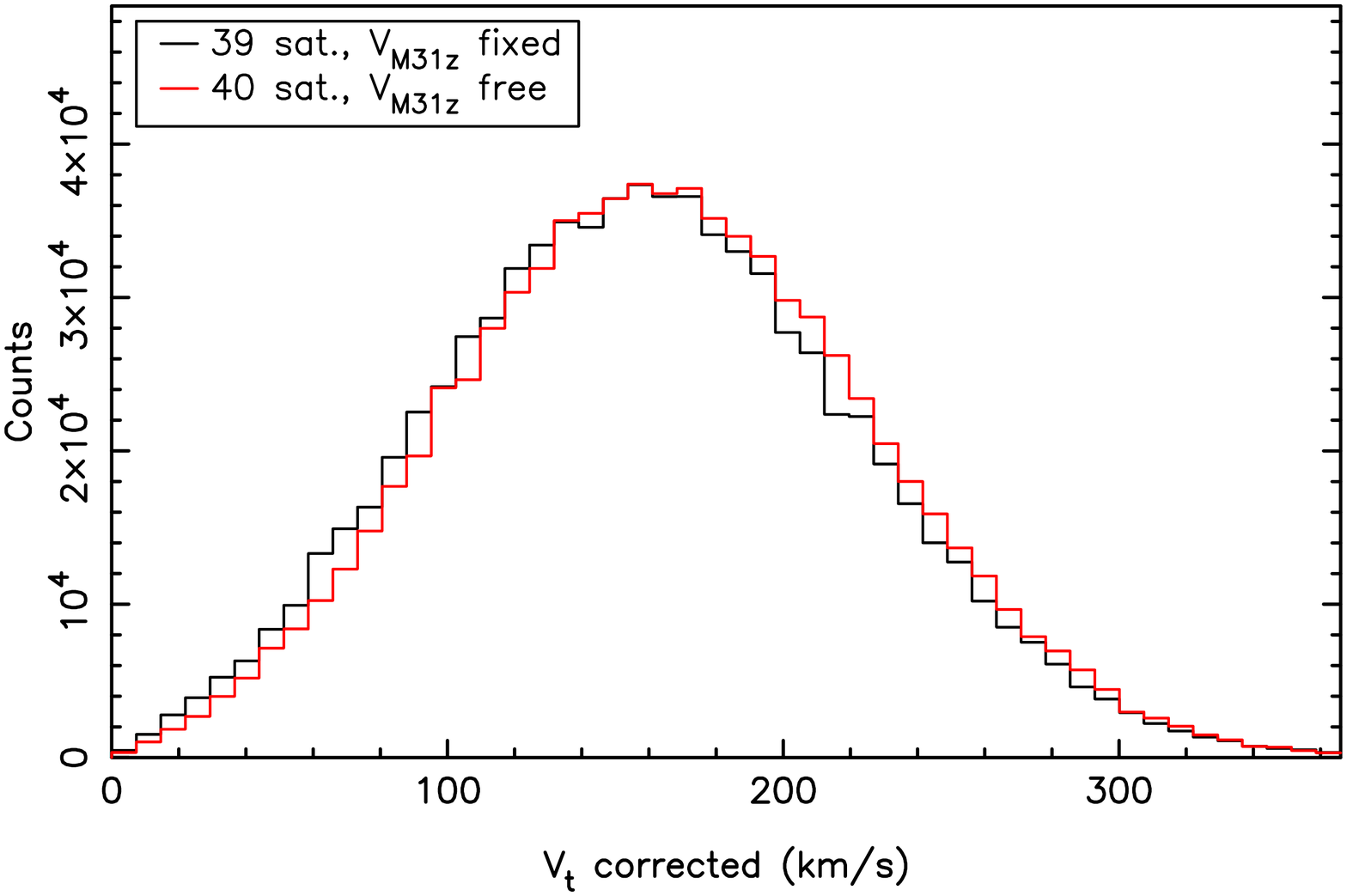} \includegraphics[scale=0.315]{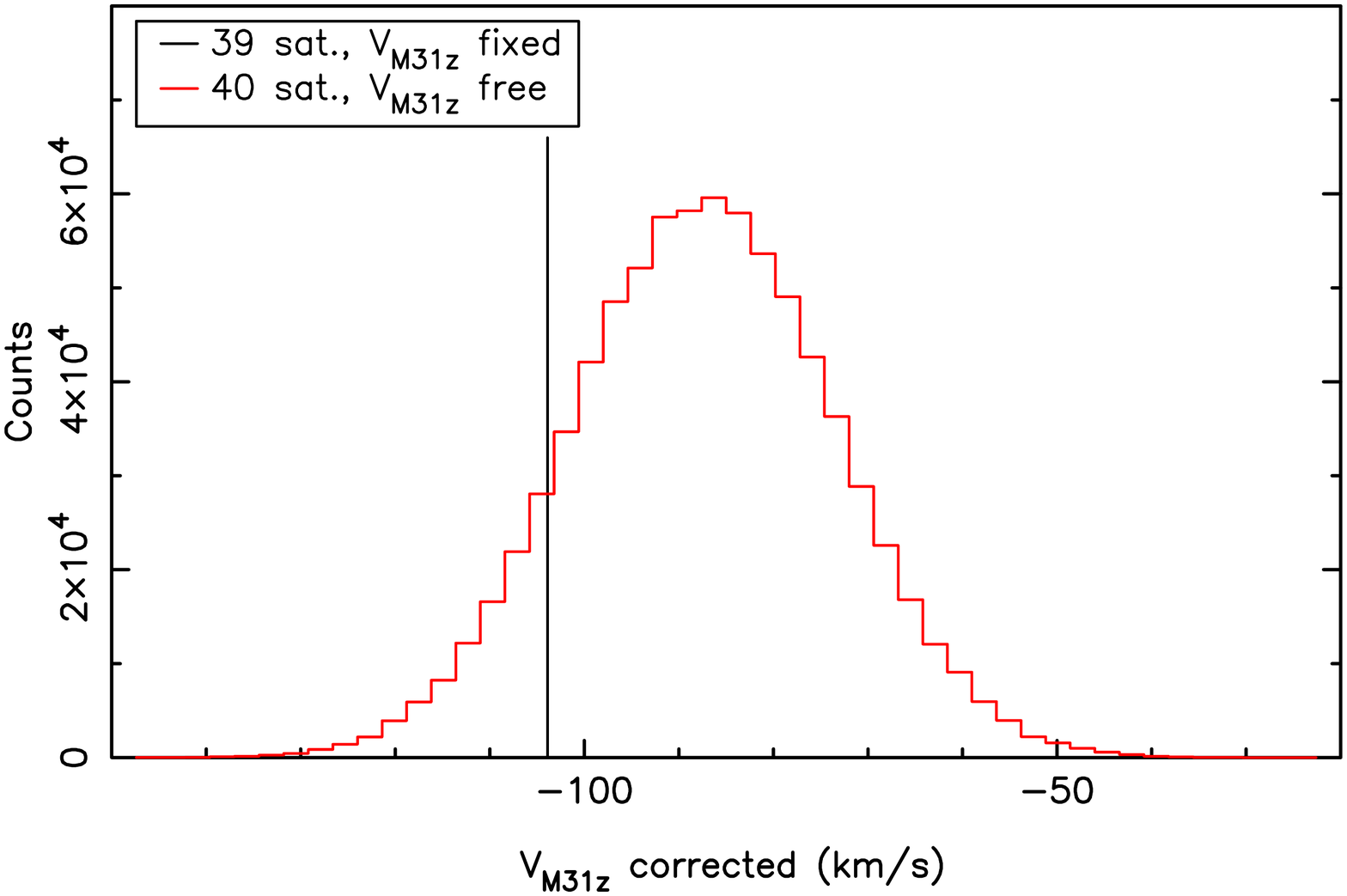}
\caption{\small\label{vpdf}
Probability density functions of the motion of the M31 system for $v_{M31x}$, $v_{M31y}$, $v_t$ and $v_{M31z}$ resulting from our MCMC method. The black line represents the PDFs for 39 satellites when $v_{M31z}$ is fixed whereas the red line traces the PDFs for 40 objects (39 satellites plus M31) with $v_{M31z}$ considered as a free parameter. In the bottom-right hand panel, the vertical line shows the fixed value of $v_{M31z}$ for the experiment with 39 satellites.}
\end{center}
\end{figure*}

The above discussion motivates us to consider the samples with 39 or 40 satellites (i.e. the samples without object rejection) as the preferred configurations; the probability density functions derived from our MCMC chains from the analyses of these samples are shown in Figure~\ref{vpdf}. The spatial distribution of corrected velocities, and velocity residuals for the full sample of 40 satellites is shown in Figure~\ref{4a}. Along the direction of motion our method infers (arrow in panel b), one can notice visually a velocity gradient by eye. Once this model for the motion is removed (panel c) no large-scale pattern in the residuals is evident. It is interesting to note from panel (c) that after accounting for the bulk motion of the system, some of the kinematic coherence of the VTP is lost: while the VTP satellites to the South of M31 predominantly have negative velocities (blue), the Northern satellites no longer have large positive velocities ({\it c.f.} \citealt{Ibata13}).

\section{Discussion and Conclusions}\label{cclpropre}

We have shown that the method we have developed allows the three-dimensional velocity of the halo to be measured accurately. Our tests using the ELVIS suite of simulations gave rise to uncertainties on the transverse motion of less than $65\kms$, irrespective of whether the halos were isolated or in Local Group type pairs. The further tests, building models similar to the observations allowed us to demonstrate that the particular sky positions that the current sample of satellites are situated at, do not give rise to significant biases in the measured proper motion. Our model also showed that an incorrect estimate of the parameters of the NFW input model (M$_{200}$, r$_{200}$, $c$ and three-dimensional position) has a negligible effect on our results. In contrast, the uncertainties on the three-dimensional motion of the Sun within the Milky Way can cause significant systematic errors. For instance, lowering the tangential motion of the observer by $35\kms$, causes the motion of the Andromeda system to fall by $15\kms$.

Due to their enormous masses, the haloes of giant galaxies should dominate their environment, so we may expect it to be natural that the baryonic disk formed at the centre of this structure. This was the basis for our first set of measurements, where we analysed the motions of the 39 satellites, assuming that the radial velocity of the halo shares the radial velocity of M31 itself. In this way the motion of the system was established.

In a second set of measurements, we dropped the requirement for the radial velocity of M31 and the larger system to be identical. M31 was then considered to be just another satellite particle within this system. By analysing several selections of satellites, we are able to demonstrate the good coherence and stability of our results. 

When placed into the Heliocentric frame, our measurements are mostly in good agreement with the earlier study of vdMG08, see Figure~\ref{resultsHelio}, based on a much smaller sample of satellites. However, we are forced to draw different conclusions to vdMG08, partially due to recent improvements in the determination of the Solar motion. The remaining differences are in the measurements themselves. In Figure~\ref{resultsHelio}, the most discrepant of the measurements by vdMG08 from our results is that due to the ``outer Local Group galaxy sample'', and it is this estimate that contributes most to the vdMG08 weighted average. However, their outer Local Group galaxy sample consists of only 5 galaxies, and given the statistical nature of their test, there are strong grounds to be concerned about the tiny sample size.
While the HST measurements shown in Figure~\ref{resultsHelio} pertain to the motion of M31 rather than that of the whole system, we note that the HST Disk field measurement is completely consistent with our findings. As noted in Section~\ref{sec:Introduction}, it is possible that the spheroid field could contain some unidentified kinematic substructure that biases that measurement, and current dynamical models of the Giant Stellar Stream may be incomplete.

Another interpretation of the kinematics of the M31 satellite system is that the observed velocity gradient reflects some intrinsic rotation of the halo. This possibility was explored recently by \cite{Deason11}, who found the need for a rotation of 
$v_{\phi}$ = (62 $\pm$ 34) ($R/10 \kpc)^{-1/4} \kms$ , where $R$ is the projected radius from the axis of rotation. It seems difficult to disentangle such a rotation from a bulk motion such as that considered in the present contribution, as the two effects will be quite similar. However, we note that in Section~\ref{sec_cosmo}, we calibrated our method against cosmological simulations that do have rotation, and found no significant bias. Moreover, in Section \ref{plan}, when we artificially imposed the presence of a plane of satellites, we found only a slight effect, contained within the range of uncertainties, on the recovered tangential velocity.

\subsection{Implications}

The inferred transverse motion reported in this contribution turns out to be surprisingly high. The value of what we consider to be the most robust velocity measurement, corrected for the Solar motion, is (-111.5 $\pm$ 70.2, 99.4 $\pm$ 60.0, -87.5 $\pm$ 13.8) $\kms$, in the East, North, and radial directions, respectively. Note that the radial velocity of the M31 system, determined in this way, lies within one sigma of the radial velocity of M31 corrected for the Solar motion (-103.9 $\pm$ 4.0 km/s). This suggests that M31 does indeed share the same kinematics as the M31 system of satellites, and the dominant dark matter halo.

The transverse motion of the complex turns out to be somewhat larger than the radial motion, with a value of 164.4 $\pm$ 61.8 $\kms$, i.e. a radial orbit with respect to the Milky Way can be rejected at $\approx 2.7\sigma$. The direction of this motion, almost normal to the Supergalactic Plane (and directed away from it), is in excellent agreement with the prediction of transverse motion by \citet{Peebles01}, based on the action principle. Using the positions and redshifts of the principal galaxies out to 20~Mpc, they predict the local distribution of mass and estimate the transverse velocity for a number of galaxies. According to their calculations, Andromeda should have a transverse velocity of $150\kms$ directed either toward or away from the Supergalactic Plane.
Given this agreement with a large transverse velocity , it will be interesting to investigate how this non-radial velocity affects the analysis of the Local Group mass, via the Timing Argument \citep[e.g.][]{Penarrubia14}.

Over time, galaxies acquire angular momentum from the various gravitational interactions they have with their neighbouring galaxies. Even if part of the torque is ``absorbed'' by the dark matter halo \citep{Barnes88}, some fraction is imparted onto the baryonic structures. Thus, following the approach pioneered by \cite{Raychaudhury89}, it is possible to estimate the magnitude of the tidal forces imparted on the Local Group over the course of its past evolution. These authors showed that the torque imparted on the Milky Way and Andromeda, caused by the action of external galaxies (within 10~Mpc), is not negligible at $z=0$. They estimated that the resulting transverse velocity of M31 should correspond to approximately $40\kms$.
According to the uncertainties, this value is 2-sigma lower than what we find in this study.
However, it shows that a transverse velocity of M31 can be expected to arise from the tidal field that the Local Group is subject to.
It is worth noting that the number, distances and masses of nearby galaxies have been significantly updated since that earlier work was published, which may be interesting to re-examine in the light of modern data.

The transverse motion of the M31 satellite system revealed by our study should also be placed in its context in terms of the dynamics of objects in their environment {\it within} the Local Group. Our result is effectively a measurement relative to the motion of the Milky Way (and within the uncertainties of the Sun's motion). So the motion of the Milky Way is implicitly subsumed within our analysis. However, recently \citet{Besla12} have suggested that the mass of the Large Magellanic Cloud (LMC) is of order 10$^{11}$ M$_{\odot}$. The LMC appears to be on its first passage around our Galaxy, on an orbit that exceeds 6~Gyr, and with a radial velocity whose present magnitude exceeds $300\kms$. Qualitatively, this means that if the total halo mass of the Milky Way lies in the vicinity of 1 $\times$ 10$^{12}$ M$_{\odot}$, our Galaxy may have accelerated up to a velocity of $\sim 60\kms$ in the direction of the LMC. This back-of-the-envelope estimate is consistent with recent findings by \citet{Gomez15}, who estimate an upper limit of this velocity of $75\kms$. In subsequent work, it will be interesting to examine the possible effect of the LMC on our measurement of the space velocity of the M31 system.

The projection onto the sky of the velocity of the M31 system that we have measured is aligned with the vector connecting M33 to M31. This orientation may not be fortuitous given that M33 is the third most massive galaxy in the Local Group, and that its accretion during the formation of the M31 system may have changed the internal dynamics of the system. The proper motion of M33 has been measured by \citet{Brunthaler05}.
Using that proper motion study, \cite{Loeb05} placed constraints on the proper motion of M31.
They found an amplitude of $100 \pm 20 \kms$, consistent with our results.
In the Heliocentric frame, we find $v_{M31x} =  33.1 {\pm70.2}\kms$, $ v_{M31y} = 19.6 {\pm60.0}\kms$, which is not in the North-west direction excluded by \cite{Loeb05}.
It will be interesting to re-explore this issue with orbital models of M33 within the M31 system using the kinematics that we have determined.
The impact of this satellite galaxy on the Andromeda system can thereby be quantified.

Finally, it will be very interesting to examine whether the transverse motion that we have detected calls into question whether the Local Group is gravitationally bound. Indeed, is it instead just a ``Local Flyby''? If we consider that the Andromeda system really does have the velocity that we have measured, simple orbital calculations (integrating the equations in \citealt{Partridge13}), show that the distance of closest approach between the two galaxies is $\approx 550\kpc$. Setting up a hydrodynamical simulation with a similar approach to \citep{Cox08} would allow the exploration of the interaction, including the investigation of the intensity of tidal forces on the two galaxies.

\section*{Acknowledgments}

R.I. would like to thank Jorge Pe\~narrubia for several very insightful conversations.

\bibliographystyle{mn2e}


\appendix

\label{lastpage}

\end{document}